# Fuzzy Keyword Search over Encrypted Data using Symbol-Based Trie-traverse Search Scheme in Cloud Computing

*P. Naga Aswani, K. Chandra Shekar*

**M. Tech Student, Associate Professor, CSE Dept.**

**puduru.aswani@gmail.com,chandhra2k7@gmail.com**

*Abstract*— As Cloud Computing becomes prevalent, more and more sensitive information are being centralized into the cloud. Although traditional searchable encryption schemes allow a user to securely search over encrypted data through keywords and selectively retrieve files of interest, these techniques support only exact keyword search. In this paper, for the first time we formalize and solve the problem of effective fuzzy keyword search over encrypted cloud data while maintaining keyword privacy. Fuzzy keyword search greatly enhances system usability by returning the matching files when users' searching inputs exactly match the predefined keywords or the closest possible matching files based on keyword similarity semantics, when exact match fails. In our solution, we exploit edit distance to quantify keywords similarity and develop two advanced techniques on constructing fuzzy keyword sets, which achieve optimized storage and representation overheads. We further propose a brand new symbol-based trie-traverse searching scheme, where a multi-way tree structure is built up using symbols transformed from the resulted fuzzy keyword sets. Through rigorous security analysis, we show that our proposed solution is secure and privacy-preserving, while correctly realizing the goal of fuzzy keyword search. Extensive experimental results demonstrate the efficiency of the proposed solution.

## 1. INTRODUCTION

As Cloud Computing becomes prevalent, more and more sensitive information are being centralized into the cloud, such as emails, personal health records, government documents, etc. By storing their data into the cloud, the data owners can be relieved from the burden of data storage and maintenance so as to enjoy the on-demand high quality data storage service.

However, the fact that data owners and cloud server are not in the same trusted domain may put the our sourced data at risk, as the cloud server may no longer be fully trusted. It follows that sensitive data usually should be encrypted prior to outsourcing for data privacy and combating unsolicited accesses.

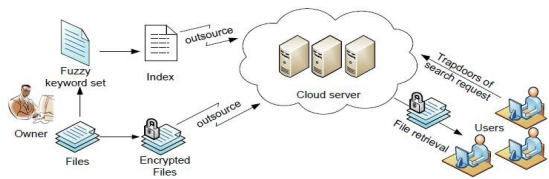

Fig. 1: Architecture of the fuzzy keyword search

However, data encryption makes effective data utilization a very challenging task given that there could be a large amount of outsourced data files. Moreover, in Cloud Computing, data owners may share their outsourced data with a large number of users. The individual users might want to only retrieve certain specific data files they are interested in during a given session. One of the most popular ways is to selectively retrieve files through keyword-based search instead of retrieving all the encrypted files back which is completely impractical in cloud computing scenarios. Such keyword-based search technique allows users to selectively retrieve files of interest and has been widely applied in plaintext search scenarios, such as Google search [1].

One of the most popular ways or techniques is to selectively retrieve files through keyword based search instead of retrieving all the encrypted files back. The data encryption also demands the protection of keyword privacy since keywords usually contain important information related to the data files. The existing searchable encryption techniques do not suit for cloud computing scenario because they support only exact keyword search. This significant drawback of existing schemes signifies the important need for new methods that support searching flexibility, tolerating both minor types and format inconsistencies. A secure fuzzy search [S. Ji, 2009] capability is demanded for achieving enhanced system usability in Cloud Computing. The main problem is how efficiently searching the data and retrieves the results in most secure and privacy preserving manner.

For retrieving the data in a most secure and privacy preserving manner the keyword searching technique is used and to search the data in more efficient manner, the fuzzy keyword search is introduced. So efficiency of fuzzy keyword search is the main aspect in the security of data retrieval. When the files are retrieved in an efficient manner, most relevant data can be retrieved. The existing system is mainly focusing on the „fuzzy keyword search" method. The data that is outsourced is encrypted, constructs fuzzy sets based on both wild card technique and gram based technique, and also introduced a symbol-based trie-traverse search scheme [Xin Zhou 2006, J. Li 2009], where a multi-way tree was constructed for storing the fuzzy keyword set and finally retrieving the data.[2]

In this paper, we focus on enabling effective yet privacy preserving fuzzy keyword search in cloud computing .to the best of our knowledge we formalize for the first time the problem of effective fuzzy keyword search over encrypted cloud data.

While maintaining keyword privacy. Fuzzy keyword search greatly enhances system usability by returning the matching files when user's searching inputs exactly match.

Pre defined keyword or the closest possible matching files based on keyword similarity semantics, when exact match fails. More specifically, we use edit distance to quantify keywords similarity and develop a novel technique, i.e., a wildcard-based technique, for the construction of fuzzy keyword sets.. Based on the constructed fuzzy keyword sets, we propose an efficient fuzzy keyword search scheme. Through rigorous security analysis, we show that the proposed solution is secure and privacy-preserving, while correctly realizing the goal of fuzzy keyword search.

## 2.RELATED WORK

**Plaintext fuzzy keyword search:** Recently, the importance of fuzzy search has received attention in the context of plain text searching in information retrieval community [11]–[13].They addressed this problem in the traditional information access paradigm by allowing user to search without using try-and-see approach for finding relevant information based on approximate string matching. At the first glance, it seems possible for one to directly apply these string matching algorithms to the context of searchable encryption by computing the trapdoors on a character base within an alphabet. However, this trivial construction suffers from the dictionary and statistics attacks and fails to achieve the search privacy.

**Searchable encryption:** Traditional searchable encryption [2]–[8], [10] has been widely studied in the context of cryptography. Among those works, most are focus on efficiency improvements and security definition formalizations. The first construction of searchable encryption was proposed by Song et al. [3], in which each word in the document is encrypted independently under a special two-layered encryption construction. [4] Proposed to use Bloom filters to construct the indexes for the data files.

To achieve more efficient search, Chang et al. [7] and Curt molaet al. [8] both proposed similar "index" approaches, where a single encrypted hash table index is built for the entire Owner outsource Files outsource Encrypted Files Trap doors of search request File retrieval Fuzzy keyword set Index Users Cloud server file collection. In the index table, each entry consists of the trap door of a keyword and an encrypted set of file identifiers whose corresponding data files contain the keyword. As a complementary approach, Boneh et al. [5] presented a public-key based searchable encryption scheme, with an analogous scenario to that of [3]. Note that all these existing schemes support only exact keyword search, and thus are not suitable for Cloud Computing. Others**.** Private matching [14], as another related notion has been studied mostly in the context of secure multiparty computation to let different parties compute some function of their own data collaboratively without revealing their data to the others. These functions could be intersection or approximate private matching of two sets, etc. The private information retrieval [15] is an often-used technique to retrieve the matching items secretly, which has been widely applied in information retrieval from database and usually incurs unexpectedly computation complexity.

**Complete Search:** Bast et al. proposed techniques to support "Complete Search," in which a user types in keywords letter by letter, and the system finds records that include these keywords (possibly at deferent places) [4, 5, 2, 3].Our work differs from theirs as follows.

(1) Complete Search mainly focused on compression of index structures, especially in disk-based settings. Our work focuses on efficient query processing using in-memory indexes in order to achieve a high interactive speed.

(2) (2) Our work allows fuzzy search, making the computation more challenging.

(3) (3) For a query with multiple keywords, Complete Search mainly caches the results of the query excluding the last keyword, which may require computing and caching a large amount of intermediate results.

## 3. PROBLEM FORMULATION

A. System Model

In this paper, we consider a cloud data system consisting of data owner, data user and cloud server. Given a collection

Of $n$ encrypted data files $C = (F1, F2, . . . , FN)$ stored in the cloud server, a predefined set of distinct keywords $W = \{w1, w2, ...,wp\}$, the cloud server provides the search service for the authorized users over the encrypted data $C$. We assume the authorization between the data owner and users is appropriately done. An authorized user types in a request to selectively retrieve data files of his/her interest. The cloud server is responsible for mapping the searching request to set of data files, where each file is indexed by a file ID and linked to a set of keywords. The fuzzy keyword search scheme returns the search results according to the following rules:

1) if the user's searching input exactly matches the pre-set keyword, the server is expected to return the files containing the keyword1

2) if there exist typos and/or format inconsistencies in the searching input, the server will return the closest possible results based on pre-specified similarity semantics (to be formally defined in section III-D). Architecture of fuzzy keyword search is shown in the Fig. 1.

### B. Threat Model

We consider a semi-trusted server. Even though data files are encrypted, the cloud server may try to derive other sensitive information from users' search requests while performing keyword-based search over $C$. Thus, the search should be conducted in a secure manner that allows data files to be securely retrieved while revealing as little information as possible to the cloud server. In this paper, when designing fuzzy keyword search scheme, we will follow the security definition deployed in the traditional searchable encryption [8]. More specifically, it is required that nothing should be leaked from the remotely stored files and index beyond the outcome and the pattern of search queries.

### C. Design Goals

In this paper, we address the problem of supporting efficient yet privacy-preserving fuzzy keyword search services over encrypted cloud data. Specifically, we have the following goals: i) to explore new mechanism for constructing storage efficient fuzzy keyword sets; ii) to design efficient and effective fuzzy search scheme based on the constructed fuzzy keyword sets; iii) to validate the security of the proposed scheme.

### D. Preliminaries

**Edit Distance** There are several methods to quantitatively measure the string similarity. In this paper, we resort to the well-studied edit distance [16] for our purpose. The edit distance ed ($w1, w2$) between two words $w1$ and $w2$ is the number of operations required to transform one of them into the other. The three primitive operations are

1) Substitution: changing one character to another in a word;

2) Deletion: deleting one character from a word;

3) Insertion: inserting a single character into a word. Given a keyword $w$, we let $S_{w,d}$ denote the set of words $w\_$ satisfying $d(w,w\_) \leq d$ for a certain integer $d$.

**Fuzzy Keyword Search** Using edit distance, the definition of fuzzy keyword search can be formulated as follows: Given a collection of $n$ encrypted data files $C = (F1, F2, . . . , FN)$ stored in the cloud server, a set of distinct keywords $W = \{w1, w2, ...,wp\}$ with predefined edit distance $d$, and a searching input $(w, k)$ with edit distance $k$ ($k \leq d$), the execution of fuzzy keyword search returns a set of file IDs whose corresponding data files possibly contain the word $w$, denoted as $FID_w$: if $w = w_i \in W$, then return $FID_{wi}$; otherwise, if $w\_ \in W$, then return $\{FID_{wi}\}$, where ed$(w,wi) \leq k$. Note that the above definition is based on the assumption that $k \leq d$. In fact, $d$ can be different for distinct keywords and the system will return $\{FID_{wi}\}$ satisfying ed$(w,wi) \leq \min\{k, d\}$ if exact match fails.

## 4. CONSTRUCTIONS OF EFFECTIVE FUZZY KEYWORD SEARCH IN CLOUD

The key idea behind our secure fuzzy keyword search is two-fold: 1) building up fuzzy keyword sets that incorporate not only the exact keywords but also the ones differing slightly due to minor typos, format inconsistencies, etc.; 2) designing an efficient and secure searching approach for file retrieval based on the resulted fuzzy keyword sets. In this section, we will focus on the first part, i.e., building storage-efficient fuzzy keyword sets to facilitate the searching process.

### 4.1 The Straightforward Approach

Before introducing our constructions of fuzzy keyword sets, we first propose a straightforward approach that achieves all the functions of fuzzy keyword search, which aims at providing an overview of how fuzzy search scheme works.

Assume $\Pi = (\text{Setup}(1^\lambda), \text{Enc}(sk, \cdot), \text{Dec}(sk, \cdot))$ is a symmetric encryption scheme, where sk is a secret key, Setup$(1^\lambda)$ is the setup algorithm with security parameter $\lambda$, Enc$(sk, \cdot)$ and Dec$(sk, \cdot)$ are the encryption and decryption algorithms, respectively. The scheme goes as follows: We can begin by constructing the fuzzy keyword set $S_{wi, d}$ for each keyword $wi \in W$ ($1 \leq i \leq p$) with edit distance d. The intuitive way to construct the fuzzy keyword set of wi is to enumerate all possible values.

2) To search with w, the authorized user computes the trapdoor Tw of w and sends it to the server; 3) Upon receiving the search request Tw , the server compares it with the index table and returns all the possible encrypted file identifiers{Enc(sk, FIDwi kwi)} according to the fuzzy keyword definition in section III-D. The user decryptsthe returned results and retrieves relevant files of interest.

This straightforward approach a p p a r e n t l y provides fuzzy keyword search over the encrypted files while achieving search privacy using the technique of secure trapdoors. However, this approach has serious efficiency disadvantages. The simple enumeration method in constructing fuzzy key- word sets would introduce large storage complexities, which greatly affect the usability. Recall that in the definition of edit distance, substitution, deletion and insertion are three kinds of operations in computation of edit distance. The numbers of all similar words of wi satisfying ed(wi , w′ ) ≤ d for d = 1, 2 and 3 are approximately $2k \times 26$, $2k^2 \times 26^2$ , and $4k^3 \times 26^3$, respectively. For example, assume there are $10^4$ keywords in the file collection with average keyword length 10 and d = 2.

The output length of hash function is 160 bits. The resulted storage cost for the index will be 30GB. Therefore, it brings forth the demand for fuzzy keyword sets with smaller size.

## 4.2 Advanced Techniques for Constructing Fuzzy Keyword Sets

To provide more practical and effective fuzzy keyword search constructions with regard to both storage and search efficiency, we now propose two advanced techniques to improve the straightforward approach for constructing the fuzzy keyword set. Without loss of generality, we will focus on the case of edit distance d = 1 to elaborate the proposed advanced techniques. For larger values of d, the reasoning is similar. Note that both techniques are carefully designed in such a way that while suppressing the fuzzy keyword set, they will not affect the search correctness, as will be described in section 5.

**Wildcard-based Fuzzy Set Construction** In the above straightforward approach, all the variants of the keywords have to be listed even if an operation is performed at the same position. Based on the above observation, we proposed to use a wildcard to denote edit operations at the same position. The wildcard-based fuzzy set of wi with edit distance d is denoted as S,· · · , S }, where S' denotes the set of words w,ti  with t wildcards. Note each wildcard represents an edit operation on wiw. The procedure for wildcard-based fuzzy set construction is shown in

Algorithm 1. For example, for the keyword CASTLE with the pre- set edit distance 1, its wildcard-based fuzzy keyword set can be constructed as S

= {CASTLE, *CASTLE, *ASTLE, C*ASTLE, C*STLE, · · · , CASTL*E, CASTL*, CASTLE*}. The total number of variants on CASTLE constructed in

this way is only 13 + 1, instead of 13 × 26 + 1 as in the above exhaustive enumeration approach

CASTLE,1 when the edit distance is set to be 1. Generally, for a given keyword w with length l, the size of Swi,1i will be only 2l + 1 + 1, as compared to (2l + 1) × 26 + 1 obtained in the straightforward Approach. The larger the pre-set edit distance, the more storage overhead can be reduced: with the same setting of the example in the straightforward approach, the proposed technique can help

Reduce the storage of the index from 30GB to approximately 40MB.

**Gram-Based Fuzzy Search:** There have been recent studies to support efficient fuzzy string search using grams [7, 1,8, 10, 15,16, 13, 12, 11, 20, 6]. A gram of a string is a substring that can be used as a signature for efficient search. These algorithms answer a fuzzy query on a collection of strings using the following observation: if a string r in the collection is similar to the query string, then should share a certain number of common grams with the query string. This "count filter" can be used to construct gram inverted lists for string ids to support efficient search. .

We evaluated some of the representative algorithms. The results showed that, not surprisingly, they are not as efficient as trie-based incremental-search algorithms, mainly because it is not easy to do incremental computation on gram lists, especially when a user types in a relatively short pre fix, and count filtering does not give enough pruning power to eliminate false positives.

```
Algorithm 1 Wildcard-based Fuzzy Set Construction
1: procedure CreateWildcardFuzzySet(w_i, d)
2:    if d > 1 then
3:        Call CreateWildcardFuzzySet(w_i, d − 1);
4:    end if
5:    if d = 0 then
6:        Set S'_{w_i,d} = {w_i};
7:    else
```

```
Algorithm 2 Gram-based Fuzzy Set Construction
1: procedure CreateGramFuzzySet(w_i, d)
2:    if d > 1 then
3:        Call CreateGramFuzzySet(w_i, d − 1);
4:    end if
5:    if d = 0 then
6:        Ŝ_{w_i,d} = {w_i};
7:    else
8:        for (k ← 1 to |S'_{w_i,d−1}|) do
9:            for j ← 1 to 2 ∗ |S'_{w_i,d−1}[k]| + 1 do
10:               Set fuzzyword as Ŝ_{w_i,d−1}[k];
11:               Delete the j-th character;
12:               if fuzzyword is not in S'_{w_i,d−1} then
13:                   Set Ŝ_{w_i,d} = Ŝ_{w_i,d} ∪ {fuzzyword}
14:               end if
15:           end for
16:       end for
17:   end if
18: end procedure
19: end procedure
```

# 5. EFFICIENT FUZZY SEARCHING SCHEMES

As shown in section 4, the size of fuzzy keyword set is greatly reduced using the proposed advanced techniques. However, the above constructions introduce another challenge: How to generate the search request and how to perform fuzzy keyword search? In the straightforward approach, because the index is created by enumerating all of fuzzy words for each keyword, there always exists matching words for the search request as long as the edit distance between them is equal or less than d. To design fuzzy search schemes based on the fuzzy keyword sets constructed from wildcard-based or gram-based technique, we compute the searching request regarding (w, k) as $\{T_{w'}\}$ $w' \in S_{w,k}$, where $S_{w,k} = \{S_{w,0}, S_{w,1}, \cdots, S_{w,k}\}$ is generated in the same way as in the fuzzy keyword set construction. In this section, we will show how to achieve fuzzy keyword search based on the fuzzy sets constructed from the proposed advanced techniques. For simplicity, we will only consider the fixed d in our scheme designs. In this section, we start with some intuitive solutions, the analysis of which will motivate us to develop more efficient ones.

## 5.1 The Intuitive Solutions

Based on the storage-efficient fuzzy keyword set constructed as above, an efficient way to realize fuzzy keyword search is to use the traditional listing approach. Specifically, the scheme goes as

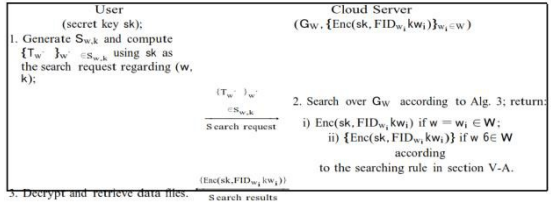

Fig. 2: Protocol for the symbol-based trie-traverse fuzzy keyword search

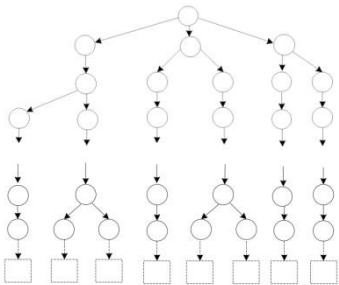

Fig. 3: An example of integrated symbol-based index for all words in the fuzzy keyword set.

## 5.2 The Symbol-based Trie-Traverse Search Scheme

To enhance the search efficiency, we now propose a symbol-based trie-traverse search scheme, where a multi-way tree is constructed for storing the fuzzy keyword set $\{S_{w_i,d}\}$ $w_i \in W$ over a finite symbol set. The key idea behind this construction is that all trapdoors sharing a common prefix may have common nodes. The root is associated with an empty set and the symbols in a trapdoor can be recovered in a search from the root to the leaf that ends the trapdoor. All fuzzy words in the trie can be found by a depth-first search. Assume $\Delta = \{a_i\}$ is a predefined symbol set, where the number of different symbols is $|\Delta| = 2n$, that is, each symbol $a \in \Delta$ can be denoted by n bits. The scheme, as described in Fig. 2, works as follows:

```
Algorithm 3 SearchingTree
1:  procedure SearchingTree({T_w})
2:      for i ← 1 to |{T_w}|
        do
3:          set currentnode as root of G_w;
4:          for j ← 1 to l/n do
5:              Set α as α_j in the i-th T;
6:              if no child of currentnode contains α then
7:                  break;
8:              end if
9:              Set currentnode as child containing α;
10:         end for
11:         if currentnode is leaf node then
12:             Append currentnode.FIDs to resultIDset;
13:             if i = 1 then
14:                 return resultIDset;
15:             end if
16:         end if
17:     end for
18:     return resultIDset;
19: end procedure
20: end procedure
```

## 5.3 Supporting Multiple Users

In this section, we consider a natural extension from the previous single-user setting to multi-user setting, where a data owner stores a file collection on the cloud server and allows an arbitrary group of users to search over his file collection. Let BE = (KeyGenBE, EncBE, DecBE) be a broadcast encryption scheme providing revocation-scheme security against a coalition of all revoked users[27]. additionally, let π be a pseudo-random permutation. The index computation is almost the same as the single-user setting except for each trapdoor $T_w$, a pseudo-random permutation $\pi(\xi, \cdot)$ is applied with a secret key $\xi$ which is encrypted with the broadcast encryption scheme and stored on the server.

To search with (w, k), an authorized user computes trapdoors {π(ξ, Tw′)}w′∈Sw,k with a secret key ξ which is distributed by the data owner.

Upon receiving the request, the server recovers the trapdoors by computing π−1 (ξ,π(ξ,Tw′)). Because the key ξ currently used is only known by the server and the set of currently authorized users, the search request is valid only if the user is not revoked. Each time a user is revoked, the data owner picks a new ξ and stores it on the server encrypted such that only non-revoked users can decrypt it. After the update, the server will use the new ξ to compute π−1 (ξ, ·) for following search requests. Furthermore, the revoked users cannot recover the current ξ and thus, their requests will not yield valid trapdoors after the server applies π−1 (ξ, ·).

# 6. VERIFIABLE FUZZY KEYWORD SEARCH SCHEME

In this section, we present the proposed scheme in details and the security analysis. Our scheme consists of five algorithms (Key Gen; Build index; trapdoor; search; verify).

## 6.1 Construction of the VFKS Scheme

*Key gen*

In this process, the user generates the index generation key and document encryption key. The Key gen is a randomized key generation algorithm, which generate the keys in this way: sk; sk0: R f0;1g:

*Build index*

In this process, we incorporate the generation method of the symbol-based index in [11] and the creation way of searching tree in [4] to construct a new symbol-based tree GW. The key idea behind this construction is that all trapdoors sharing a common prefix may have common nodes. The root is associated with an empty set and the symbols in a trapdoor can be recovered in a search from the root to the leaf that ends the trapdoor [11]. All fuzzy keywords in the tire can be found by a depth-first search. Assume D = faig is a pre defined symbol set, where the number of different symbols is jDj = 2n, that is each symbol ai 2 D can be denoted by n bits. The algorithm works as follows:

(1) Initialization: – The user scan the D and build W, the set of distinct keywords of D.– The user outsource the encryption document collection D to the server and receive the identifiers of each document(denote as IDfFig ). For all document of containing the keyword wi, denote the identifier set as IDwi = IDfF1gkIDfF2g:::;kID

(2) Build fuzzy set trapdoor– For each keyword wi 2W, construct the fuzzy keyword set Swi;d with the wildcard-based method g.

(3) Build symbol-based index tree: – Create a depth of l=n full 2n-binary tree GW, where each node contains two attributes (r0;r1) = (null; null) in detail and l is the out length of hash function f(x).

– For each fuzzy keyword w0i 2 Sw i, d, divided Tinto l=n parts, each n-bits hash value represents a symbol in 4. Putting all the sequence of symbol filing into the G Wand appending the corresponding identifier IDwikgk (IDwi ) to the leaf node.

Search

– Upon receiving the search request, the server divides each Tw0 into a sequence of symbols;

– Performs the search over GW using algorithm described in Fig 2 and returns the ID wi and proof to the user.

– According to the identifier, the user can get the interest documents.

Verify

– check whether the number of received proof is equals to the number of sent trapdoor.

– input the Tw and the corresponding proof, according to the algorithm test whether the server is honest.

## 6.2 Security analysis

**Data Privacy**: In this work, we consider only search privacy; because of privacy of the documents can be ensured by the encryption algorithm. That is, we focus on the confidentiality of the search request and the index T. Using the trapdoor technology, the attacker directly to get the plaintext is impossible from the cipher text. So we mostly concern the confidentiality of the index T. In [7], Li et al. has proved that the fuzzy keyword search scheme is secure regarding the search privacy. Similar arguments will work for our scheme.

**Verifiable Searchability:** we assume k steps are performed by the server. If "No" returned, we would know that the first k 1 characters are matched while Tw [k] is mismatched, which could be described by a k bit binary sequence b = (1;::::;1;0); if "Yes" is returned, b = (1;::::;1;1).

In our scheme, firstly, we check the number of proof whether is equal to the sending trapdoors. If not, we can say the server is not making a full search. If pass, we exploit a random sampling method to check. For each of Proof to be tested, As [8], Similarly, starting from the last (or k-th) step, if "Yes", verify checks the integrity of the concatenation of the document identifiers by computing a keyed hash of it and comparing with the received one. In fact, the completeness of the search outcome is examine adhere. If the server returns a fraction of the search outcome, the user can find the server is not honest. Then we test that whether the trapdoor equals the received symbol string of the proof. After that, j decreased by one. If "No", the above step is skipped. Next, verify validates the correctness of the search outcome by decrypting the first part of Tj qj[r1] = Enc(Tj;qj[r0]; parent (Tj;qj)[r1]) to get (x; y) and testing whether: (1) Tw[ j] equals x (2) Tw[ j 1] equals y. To cheat the search results, the server need to forge the proof. There are two possible case: (1) the server honestly search for a fraction of trapdoors and forge the proof or other trapdoors, at worst, the server do not any search; (2) the server forge the proof according to the received trapdoor. For the case (1), the server must can generate a valid r1 with a different memˆ =6 mem, consider the adversary do not know the sk, it can be seen a random oracle. In case (2), the adversary may use the r1 of another node, note that each node has a global unique r1, which will be rejected by verify.

In addition, the argument above can be applied recursively to the (j 1)th step in verify and so on.

## 7. PERFORMANCE ANALYSIS

We conducted a thorough experimental evaluation of the proposed techniques on real data set: the recent ten years 'IEEEINFOCOM publications. The data set includes about 2,600 publications. We extract the words in the paper titles to construct the core keyword set in our experiment. The total number of keywords is3, 262 and their average word length is7.44. Our experiment is conducted on a Linux machine with an Intel Core 2processor running at 1.86GHz and 2G DDR2-800 memory. The performance of our scheme is evaluated regarding the time cost of fuzzy set construction, the time and storage cost of index construction, the search time of the listing approach and the symbol-based trie-traverse approach.

### 7.1 Performance of Fuzzy Keyword set Construction

In section 4, we propose two advanced techniques for the construction of fuzzy keyword sets, which both can be employed in our proposed fuzzy search schemes. In our experiment, we only focus on the wildcard-based fuzzy set construction because it provides the sound results compared to the gram-based fuzzy set construction as discussed insection6.Fig.4showsthe fuzzy set construction time by using the wildcard-based approach with edit distanced=1and2.We can see that in both cases, the wildcard-based approach is very efficient and the construction time increases linearly with the number of keywords. The cost of constructing fuzzy keyword set under d=1 is much less than the case of d=2 due to the smaller set of possible wildcard-based words.

### 7.2 Performance of Fuzzy Keyword Search

Efficiency of Index Construction Given the fuzzy keyword set constructed using wildcard- based technique, we measure the time cost of index construction for the listing approach and symbol-based trie-based approach. In our experiment, we chosen d=4 and use SHA-1as our hash function with output length of l=160bits.The resulted height of the searching tree is l/n=40.

The index construction time for edit distance d=1 and d=2. Similar to the fuzzy keyword set construction, the index construction time also increases linearly with the number of distinct keywords. Compared to the listing approach, the index construction of the trie-traverse approach includes the process of building the searching tree additionally, thus its time cost is larger than that of listing approach. However, the whole index construction process is conducted off-line, thus it will not affect the searching efficiency. Table 1shows the index storage cost of the two approaches. The symbol-based trie-traverse approach consumes more storage space than the listing approach due to its multi-way tree structure. This additional storage cost, however, is not a main issue in our setting, as such index information only take up a small amount of storage space on the cloud server.

## 8. CONCLUSION

In this paper, for the first time we formalize and solve the problem of supporting efficient yet privacy-preserving fuzzy search for achieving effective utilization of remotely stored encrypted data in Cloud Computing. We design two advanced techniques (i.e., wildcard-based and gram- based techniques) to construct the storage-efficient fuzzy keyword sets by exploiting two significant observations on the similarity metric of edit distance. Based on the constructed fuzzy keyword sets, we further propose a brand new symbol-based trie-traverse searching scheme, where a multi-way tree structure is built up using symbols transformed from the resulted fuzzy keyword sets. Through rigorous security analysis, we show that our proposed solution is secure and privacy- preserving, while correctly realizing the goal of fuzzy keyword search. Extensive experimental results demonstrate the efficiency of our solution.

As our ongoing work, we will continue to research on security mechanisms that support

1) Search semantics that takes into consideration conjunction of keywords, sequence of keywords, and even the complex natural language semantics to produce highly relevant search results. And

2) Search ranking that sorts the searching results according to the relevance criteria.
3)

## Biographies

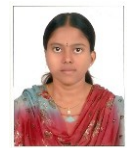

Ms. P.Naga Aswani had received her B.Tech degree in Computer Science Information Technology from Vaagdevi Institute of Technology & Science, Proddatur, JNTU Anantapur University and Pursuing M.Tech in Computer Science (Software Engineering) from Aurora's Technological And Research Institute, JNTU Hyderabad.

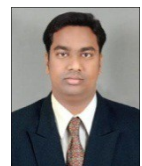

Mr. K. Chandra Shekar working as Associate Professor in the Department of Computer Science and Engineering in Aurora's Technological and Research Institute with a teaching experience of 9 Year and worked as Project Trainee. He had received his Master Degree in Software Engineering. His areas of interest include Data mining and Information Security and his Specialization in Associative Classification Mining in Non-Binary Data.